# Quantum signature in classical electrodynamics of the free radiation field

Michele Marrocco

*ENEA*

*(Italian National Agency for New Technologies, Energies and Sustainable Economic Development)*

*via Anguillarese 301, Rome, 00123 Italy*

email: michele.marrocco@enea.it

## Abstract

Quantum optics is a field of research based on the quantum theory of light. Here, we show that the classical theory of light can be equally effective in explaining a cornerstone of quantum optics: the quantization of the free radiation field. The quantization lies at the heart of quantum optics and has never been obtained classically. Instead, we find it by taking into account the degeneracy of the spherical harmonics that appear in multipole terms of the ordinary Maxwell's theory of the free electromagnetic field. In this context, the number of energy quanta is determined by a finite countable set of spherical harmonics of higher order than the fundamental (monopole). This one plays, instead, the role of the electromagnetic vacuum that, contrary to the common view, has its place in the classical theory of light.

1. **Introduction**

One of the first topics that opens our minds to the fascinating quantum reality of the physical world is the quantization of the free electromagnetic field [1-7]. The procedure is very popular and starts off by expanding the classical field in terms of plane waves, each of them associated with a particular mode of oscillation of the field. In the second part, the procedure requires the connection of the classical field modes to quantum harmonic oscillators and, by doing that, the outcome is that the electromagnetic energy is made of a countable number $n$ of energy shots (photons) plus a vacuum contribution made of half a shot (half-photon or zero-point energy). In short, the quantum-mechanical energy of a single-mode electromagnetic field made of number states is often written as follows (see, for example, pag. 185 in Ref. 5)

$$H_n = (n+1/2)\hbar\omega \qquad \text{where } n = 0, 1, 2....  \qquad (1)$$

where $\hbar$ is the reduced Planck constant and $\omega$ is the angular frequency (we use the symbol $H_n$ for the energy to distinguish it from the more common symbol $E$ that will be used later for the electric field). Eq. (1) is fundamental to many aspects of quantum optics and continues to stimulate interesting ideas with special regard to the role of the vacuum field [1, 4, 5].

In this work, we want to describe a procedure of quantization that follows a classical route to Eq. (1). The foregoing sentence might sound absurd to many because it is in marked contrast to what has been believed for decades. Indeed, it is common knowledge that the quantization is an impossible task within the classical theory of light and it comes as no surprise to read that "the energy eigenvalues are discrete, in contrast to classical electromagnetic theory" (pag. 10 in Ref. 4), or "zero-point energy... has no analogue in the classical theory" (pag. 143 of Ref. 5). Whether this is the ultimate truth is open to debate according to the results of the current work inspired by the classical multipole expansion of the electromagnetic field [8].

The key point of our argument is that the technique of multipole expansion is built on the so-called Helmholtz equation. Its solution determines the spatial dependences, which are separable on the three-dimensional sphere characterized by the spherical polar coordinates $r$, $\vartheta$ and $\varphi$. By using this and by relating the number of energy quanta to the number of spherical harmonics that appear in the angular solution of the Helmholtz equation, we are able to quantize the energy.

The demonstration is organized as follows. First of all, a short summary of the traditional quantization procedure is given. The main aspects that are important to the argument developed here are briefly reviewed, but we avoid the repetition of the ordinary treatment of the quantum harmonic oscillator that the Reader can find in several of the available sources on the subject [3-5]. Another Section reports, instead, the main calculation of the classical method that leads to the quantization of the electromagnetic energy in a number $n+1/2$ of energy parcels. Finally, in two additional Sections, we present the calculation that completes the procedure.

## 2. Summary of the standard quantization procedure of the free electromagnetic field

According to classical theory, the energy $H$ of the free-space electromagnetic field contained in a volume $V$ is calculated by means of a three-dimensional spatial integration of two contributions. One for the electric field $\mathbf{E}(\mathbf{r},t)$ and one for the magnetic field $\mathbf{B}(\mathbf{r},t)$, that is,

$$H = \frac{1}{2}\varepsilon_0 \int d\mathbf{r} |\mathbf{E}(\mathbf{r},t)|^2 + \frac{1}{2\mu_0} \int d\mathbf{r} |\mathbf{B}(\mathbf{r},t)|^2 \qquad (2)$$

where the electric permittivity and magnetic permeability of free space are related to the speed of light $c = 1/\sqrt{\varepsilon_0 \mu_0}$. In general, we cannot see why Eq. (2) should be quantized if we look naively at the classical fields $\mathbf{E}(\mathbf{r},t)$ and $\mathbf{B}(\mathbf{r},t)$. However, the energy quantization proceeds by the replacement of the electric and magnetic fields by their vector potential $\mathbf{A}(\mathbf{r},t)$. This one is chosen under the Coulomb gauge ($\nabla \cdot \mathbf{A}(\mathbf{r},t) = 0$) and decomposed as a sum of contributions from the modes $(\mathbf{k}, s)$ defined by the wave vector $\mathbf{k}$ and the state $s$ of polarization

$$\mathbf{A}(\mathbf{r},t) = \sum_{\mathbf{k},s} A_{\mathbf{k},s}(\mathbf{r},t) \, \mathbf{e}_{\mathbf{k},s} \qquad (3)$$

with $\mathbf{e}_{\mathbf{k},s}$ indicating the normalized transverse polarization vectors. The field projections $A_{\mathbf{k},s}(\mathbf{r},t)$ onto the polarization vectors are plane waves

$$A_{\mathbf{k},s}(\mathbf{r},t) = A_{\mathbf{k},s}(t) e^{i\mathbf{k}\cdot\mathbf{r}} + A^*_{\mathbf{k},s}(t) e^{-i\mathbf{k}\cdot\mathbf{r}} \qquad (4)$$

and they can be further elaborated by considering that the vector potential obeys the wave equation written for each mode of angular frequency $\omega_k$ (with $k = |\mathbf{k}|$). Finally, we find

$$A_{\mathbf{k},s}(\mathbf{r},t) = A^0_{\mathbf{k},s} e^{i\mathbf{k}\cdot\mathbf{r}-i\omega_k t} + A^{0*}_{\mathbf{k},s} e^{-i\mathbf{k}\cdot\mathbf{r}+i\omega_k t} \qquad (5)$$

and the vector potential $A_{\mathbf{k},s}(\mathbf{r},t)$ is now prepared for the purpose of calculating Eq. (2). The result is time independent and is

$$H = 2\varepsilon_0 V \sum_{\mathbf{k},s} \omega_k^2 \left| A_{\mathbf{k},s}^0 \right|^2 \tag{6}$$

At this stage, the decisive twist is the introduction of the quantum harmonic oscillator characterizing each mode $(\mathbf{k},s)$ of the vector-potential amplitudes $A_{\mathbf{k},s}^0$. The crucial correspondence presents itself in the comparison of the classical energy of Eq. (6) with the quantum-mechanical Hamiltonian of a series of independent quantum harmonic oscillators

$$\hat{H} = \sum_{\mathbf{k},s} \frac{1}{2}\hbar\omega_k \left( \hat{a}_{\mathbf{k},s} \hat{a}_{\mathbf{k},s}^+ + \hat{a}_{\mathbf{k},s}^+ \hat{a}_{\mathbf{k},s} \right) \tag{7}$$

where $\hat{a}_{\mathbf{k},s}^+$ and $\hat{a}_{\mathbf{k},s}$ are, respectively, the creation and destruction operators (for a brief account of the quantum harmonic oscillator, see Refs. 1-5). In this manner, the following correspondence can be set up

$$A_{\mathbf{k},s}^0 \rightarrow \left( \frac{\hbar}{2\varepsilon_0 V \omega_k} \right)^{1/2} \hat{a}_{\mathbf{k},s} \qquad A_{\mathbf{k},s}^{0*} \rightarrow \left( \frac{\hbar}{2\varepsilon_0 V \omega_k} \right)^{1/2} \hat{a}_{\mathbf{k},s}^+ \tag{8}$$

and everything falls into place. The bottom line is that the Hamiltonian of Eq. (7) can be changed into

$$\hat{H} = \sum_{\mathbf{k},s} \hbar\omega_k \left( \hat{a}^+_{\mathbf{k},s} \hat{a}_{\mathbf{k},s} + \frac{1}{2} \right) \quad (9)$$

whose eigenvalue is $\hbar\omega_k (n_{\mathbf{k},s} + 1/2)$ on condition that the field has an integer number $n_{\mathbf{k},s}$ of photons (photon-number state) for each mode $(\mathbf{k},s)$.

In concluding this section, let us remark that the zero-point energy corresponding to the vacuum contribution of $\hbar\omega_k /2$ is rather strange within the conceptual framework of quantum mechanics where a single photon is an indivisible unit of energy and never splits into two halves. This aspect has accompanied the development of quantum theory and remains of great attraction [1].

## 3. Quantization in classical electrodynamics

The classical quantization procedure is introduced in a slightly different manner. Instead of Eq. (2), we handle the corresponding cycle-averaged energy

$$<H> = \frac{1}{2}\varepsilon_0 \int d\mathbf{r} <|\mathbf{E}(\mathbf{r},t)|^2> + \frac{1}{2\mu_0} \int d\mathbf{r} <|\mathbf{B}(\mathbf{r},t)|^2> \quad (10)$$

The classical cycle average of Eq. (10) is indispensable for two main reasons. On the one hand, the energy does not depend on time and thus we need to eliminate the time variable. On the other, a practical reason forces us to deal with cycle-averaged energies. A detector cannot follow the fast time oscillation of the electromagnetic field. We can only measure cycle-averaged observables and, based on this, we have to deal with $<H>$ rather than $H$ of Eq. (2). But, in place of Eq. (10), we are going to work on the simpler expression

$$<H> = \varepsilon_0 \int d\mathbf{r} < |\mathbf{E}(\mathbf{r},t)|^2 > \tag{11}$$

that involves the known balance between the electric term and its magnetic counterpart. Needless to say, the procedure works even though the two terms are separated.

Next, we introduce the mode summation

$$\mathbf{E}(\mathbf{r},t) = \sum_{\mathbf{k}} \mathbf{E}_{\mathbf{k}}(\mathbf{r},t) = \sum_{\mathbf{k},s} E_{\mathbf{k},s}(\mathbf{r},t) \mathbf{e}_{\mathbf{k},s} \quad \text{with} \quad \mathbf{E}_{\mathbf{k}}(\mathbf{r},t) = \sum_{s} E_{\mathbf{k},s}(\mathbf{r},t) \mathbf{e}_{\mathbf{k},s} \tag{12}$$

plus the fact that we can decompose Eq. (11) into mode contributions

$$<H> = \varepsilon_0 \sum_{\mathbf{k},s} \int d\mathbf{r} < |E_{\mathbf{k},s}(\mathbf{r},t)|^2 > \tag{13}$$

where the components $E_{\mathbf{k},s}(\mathbf{r},t)$ of the electric field satisfy their own wave equation

$$\nabla^2 E_{\mathbf{k},s}(\mathbf{r},t) = \frac{1}{c^2} \frac{\partial^2 E_{\mathbf{k},s}(\mathbf{r},t)}{\partial t^2} \tag{14}$$

The scheme behind Eqs. (11)-(14) conveys a precise meaning. Unlike the quantum-mechanical approach outlined before, we prefer to handle the true observable, that is, the electric field rather than the vector potential that is never measured. In addition, we avoid the plane-wave expansion of Eqs. (3) and (4). This choice reduces the whole problem to a specific form of wave propagation. Our choice is different and explained below. Nonetheless, we keep the fact that the double time derivative transforms the right-hand side of Eq. (14) into $-k^2 E_{\mathbf{k},s}(\mathbf{r},t)$, having assumed that $\omega_k = ck$. With these preliminaries, we are prepared for the procedure that does not require quantum-mechanical operators.

The procedure is not new. It is partially laid out in the classical multipole expansion of electromagnetic fields [pag. 429 of Ref. 8], according to which the spatial dependences are encoded in the well-known Helmholtz equation

$$\nabla^2 E_{\mathbf{k},s}(\mathbf{r},t) + k^2 E_{\mathbf{k},s}(\mathbf{r},t) = 0 \tag{15}$$

whose use is very popular in mathematics and physics.

The solution of Eq. (15) is achieved by introducing polar coordinates and depends on the spherical Bessel and Neumann functions that define the radial component [pag. 425 of Ref. 8]. The angular dependence is instead described by the spherical harmonics that form an orthonormal basis in the Hilbert space of square-integrable functions [pag. 108 of Ref. 8]. These special functions appear in classical electrodynamics [8], classical optics [9], acoustics [10], geophysics [11], and beyond these examples of classical physics, they are central to the quantum-mechanical determination of orbital angular momenta (see Refs. 3, pp. 519-523). Soon we will discover that the spherical harmonics are central to this demonstration too.

Having made this crucial premise, we now set out the calculation of Eq. (13) for one multipole solution of the Helmholtz equation that satisfies the boundary conditions (i.e., pure spherical wave at large radial distances). The solution is chosen regular at the origin (we are in the free space, i.e, without singularities!) and, for this reason, written in dependence on the generic order $n$ of the spherical Bessel function $j_n(kr)$

$$E_{\mathbf{k},s}(r,\vartheta,\varphi,t) = E_{0,s}(t)\, j_n(kr) \sum_{m=-n}^{n} Y_n^m(\vartheta,\varphi) \; . \tag{16}$$

with the $k$ dependence incorporated in the radial component only. The amplitude $E_{0,s}(t)$ is time dependent and incorporates a phase factor that determines the oscillation direction of the field lying in the transverse plane defined by the polarization unit vectors $\mathbf{e}_{\mathbf{k},s}$ of Eq. (12). However, regardless of these secondary details, we turn our attention to the integration of the spherical harmonics. These make the calculation of Eq. (13) very simple. In effect, the mode energy extracted from Eq. (13) reduces to the following evaluation

$$\varepsilon_0 \int d\mathbf{r} < |E_{\mathbf{k},s}(\mathbf{r},t)|^2 > = \varepsilon_0 \int\int\int dr\, d\vartheta\, d\varphi\, r^2 \sin\vartheta < |E_{\mathbf{k},s}(r,\vartheta,\varphi,t)|^2 > \tag{17}$$

and the integration over the solid angle results in

$$\sum_{m,m'=-n}^{n} \int\int d\vartheta\, d\varphi\, \sin\vartheta\, Y_n^{m'*}(\vartheta,\varphi)\, Y_n^m(\vartheta,\varphi) = 2n+1 \tag{18}$$

that brings out a compelling result on the right-hand side. The orthonormality condition for $Y_n^m(\vartheta,\varphi)$ (see Eq. 3.55, page 108, in Ref. 8) produces a degeneracy of $2n+1$ terms in the calculation of Eq. (17). This result has to be combined with the cycle average of the square of the time dependent amplitude

$$<|E_{0,s}(t)|^2> = \frac{|E_0|^2}{2} \tag{19}$$

that generates a time-independent and phase-independent factor $|E_0|^2$ and a factor of $1/2$ [coming from the cycle average of $sin^2(\omega t)$ or $cos^2(\omega t)$ terms]. In conclusion, our calculation of the mode energy is

$$\varepsilon_0 \int d\mathbf{r} < \left|E_{\mathbf{k},s}(\mathbf{r},t)\right|^2 > = \varepsilon_0 |E_0|^2 R_n \left(n + \frac{1}{2}\right) \qquad (20)$$

with the radial integral $R_n$ written as

$$R_n = \int_0^R dr\, r^2\, j_n^2(kr). \qquad (21)$$

where $R$ is the radius of the quantization volume $V$ that we have already met in the classical-quantum correspondence of Eq. (8). It is undeniable that Eq. (20) is suggestive of something that looks like the quantum-mechanical result of Eq. (1), apart from the radial integral whose calculation will be postponed to the next Section. Here, we anticipate that its dependence on $n$ is very weak and disappears when we consider the limit of $kR >> 1$, which is reasonable because $R$ is much larger than $1/k$ for all the cases of practical interest. Therefore, if $R_n$ is independent from $n$, Eq. (20) seems to produce the quantization rule of Eq. (1). What is more, we can attribute a precise meaning to the number $n$ of energy quanta. This number represents the integer that measures the level of energy excitation delivered by the spherical harmonics of opposite secondary index $m$ and, following this line of thought, the combination of Eqs. (18), (19) and (20) suggests that a classical procedure of quantization exists with the additional picture of what the vacuum field might be. It springs from the condition $n = 0$ that leaves room for the fundamental ($m = 0$) spherical harmonic only (monopole term). Its complete spherical symmetry supports the view that the vacuum field is spatially isotropic and, by looking at Eq. (18), its contribution to the energy counts for one energy quantum. The fractional value of $1/2$ in Eq. (20) is, indeed, accidental because its appearance is caused by the cycle average of Eq. (19) and this explanation seems to be more complete than the mysterious energy splitting suggested by the quantum-mechanical understanding of the vacuum energy in Eq. (9). Furthermore, Eq. (20) identifies a peculiar value of reference for the elementary electric field amplitude $E_0$ associated with each energy quantum.

The final image we have from what has been accomplished so far is rather promising and it might suffice to unravel interesting ways of treating quantum results by means of classical concepts that are rooted in the multifaceted aspects of the Maxwell theory of the electromagnetic field [8]. Nonetheless, we proceed with the calculation even though Eq. (20) might be enough to show an alternative route to energy quantization.

**4. The radial integral and final calculation**

The method outlined in the previous section is very simple and hinges on the solution of the Helmholtz equation for the electric field in place of plane-wave expansions of the vector potential. Although this solution allows us to replicate the quantization rule of Eq. (1), the problematic role of the radial integral $R_n$ [Eq. (21)] has only been mentioned. Here, we discuss it along with its solution.

The radial part of Eq. (16) is made of multipole contributions that multiply spherical waves. It means that, when $R_n$ is calculated, we obtain a term that depends linearly on the radius $R$ plus oscillating terms. The result can be written in closed form and depends on a combination of ordinary Bessel functions. However, we are interested in the limit of $kR \gg 1$.

Now, the great value of $kR$ suppresses the oscillating terms and we find that $R_n$ is independent from $n$. In this case, being $R_n = R/(2k^2)$, Eq. (20) becomes

$$\varepsilon_0 \int d\mathbf{r} <|E_{\mathbf{k},s}(\mathbf{r},t)|^2> = \left(n + \frac{1}{2}\right)\varepsilon_0|E_0|^2 \frac{R}{2k^2}. \tag{22}$$

Regrettably, the result of the radial integral does not make Eq. (22) conform with our hopes. Its effect on the calculation contrasts with the linearity of Eq. (1). As a matter of fact, considering that $\omega_k = ck$, the result of Eq. (22) indicates a different dependence on the angular frequency. While the quantum-mechanical result is linear, we find an inverse square law that corresponds to what we know of the multipole expansion of the electromagnetic field [see page 433, Eq. (9.136), of Ref. 8]. Nonetheless, the disagreement is only apparent and more insight is needed to reconcile Eq. (22) with the quantum-mechanical result of Eq. (1).

Let us recall that we are searching for the energy associated with a well defined angular frequency $\omega$ and this objective has not yet been put into perspective. The integral of Eq. (22) is relative to a particular mode $(\mathbf{k}, s)$ and now, to meet the objective, we have to evaluate the mode summation of Eq. (13). What this amounts to is that we need to covert the summation to integration according to the usual procedure [for example, page 7 in Ref. 5]

$$\sum_{\mathbf{k},s} = 2\sum_{\mathbf{k}} \quad \rightarrow \quad \frac{V}{\pi^2}\int k^2 dk \tag{23}$$

and our final result will depend on the domain of integration. To determine it, we recall that the energy is defined in a relative way. Indeed, in quantum mechanics, we deal with quantum leaps of energy and in this sense we have to understand the quantum-mechanical energy of Eq. (1). This means, in turn, that the domain of integration of Eq. (23) has to be taken between a generic reference value $k_0$ and $k_0 + \omega/c$. The interval $[k_0, k_0 + \omega/c]$ guarantees that we are considering the right correspondence with the quantum-mechanical concept of a quantum energy of $\hbar\omega(n+1/2)$. Surprisingly, the classical result does not depend on the reference value of $k_0$ when we apply the integration of Eq. (23) and the final result is

$$<H> = \beta\left(n+\frac{1}{2}\right)\omega \qquad \text{with } \beta = \frac{\varepsilon_0}{2\pi^2 c} RV|E_0|^2 \qquad (24)$$

It is now easy to observe that Eq. (24) is nothing but Eq. (1) except for the parameters collected in $\beta$. We might also dream of jumping to the conclusion that $\beta$ equals the reduced Planck constant $\hbar$. This would stretch our reasoning to the astounding limit of a complete agreement of Eq. (24) with Eq. (1). However, we take this conclusion cautiously because we have to make sure that $\beta$ is a parameter undergoing a conservation law.

## 5. Conservation of $\beta$

The conservation of $\beta$ might be immediately deduced from the energy conservation applied to Eq. (24). Being the number $n$ of energy quanta and the angular frequency $\omega$ two given variables of constant values (the field is non-interacting), the energy is conserved and one can draw the conclusion that $\beta$ is conserved. Despite this, the conservation of $\beta$ can also be demonstrated by looking at the intrinsic angular momentum that in classical electrodynamics is defined as [pag. 489 in Ref. 2]

$$\mathbf{J}_s = \varepsilon_0 \int d\mathbf{r}\, \mathbf{E}(\mathbf{r},t) \times \mathbf{A}(\mathbf{r},t) \tag{25}$$

where, as usual, $\mathbf{A}(\mathbf{r},t)$ is the vector potential. Let us first verify that $\mathbf{J}_s$ is a constant of the motion for the free electromagnetic field. To this end, it suffices to consider the time derivative

$$\frac{d\mathbf{J}_s}{dt} = \varepsilon_0 \int d\mathbf{r} \left[ \frac{\partial \mathbf{E}(\mathbf{r},t)}{\partial t} \times \mathbf{A}(\mathbf{r},t) + \mathbf{E}(\mathbf{r},t) \times \frac{\partial \mathbf{A}(\mathbf{r},t)}{\partial t} \right] \tag{26}$$

Under the Coulomb gauge, both fields $\mathbf{E}(\mathbf{r},t)$ and $\mathbf{A}(\mathbf{r},t)$ are transverse and satisfy the wave equation so that one finds after some algebra

$$\frac{d\mathbf{J}_s}{dt} = -\varepsilon_0 \int d\mathbf{r} \left[ -\omega^2 \mathbf{A}(\mathbf{r},t) \times \mathbf{A}(\mathbf{r},t) + \frac{\partial \mathbf{A}(\mathbf{r},t)}{\partial t} \times \frac{\partial \mathbf{A}(\mathbf{r},t)}{\partial t} \right] = 0 \tag{27}$$

where the right-hand side vanishes because of the cross product properties. The conservation law of Eq. (27) is very general and holds good for any harmonically varying field that solves the wave equation and is related to the vector potential through the time derivative. We will use this fact later.

Another important feature of $\mathbf{J}_s$ is that its cycle average satisfies the following equation

$$<\mathbf{J}_s> = \frac{1}{2}\varepsilon_0 \int d\mathbf{r}\, \mathrm{Re}\left[ \mathbf{E}(\mathbf{r}) \times \mathbf{A}^*(\mathbf{r}) \right] \tag{28}$$

where $<\mathbf{J}_S>$ continues to be a constant of the motion expressed in terms of the spatially dependent fields $\mathbf{E}(\mathbf{r})$ and $\mathbf{A}^*(\mathbf{r})$. However, we are going to use an equivalent version of Eq. (28), that is

$$<\mathbf{J}_S> = -\frac{\varepsilon_0}{2\omega}\int d\mathbf{r}\,\text{Im}\left[\mathbf{E}(\mathbf{r})\times\mathbf{E}^*(\mathbf{r})\right] \qquad (29)$$

and, considering circular polarization vectors $\mathbf{e}_{\mathbf{k},\lambda}$ (with $\lambda = \pm 1$) that form a new orthonormal basis [pag. 490 in Ref. 2]

$$\mathbf{e}_{\mathbf{k},+1} = \frac{1}{\sqrt{2}}\left(\mathbf{e}_{\mathbf{k},1} + i\mathbf{e}_{\mathbf{k},2}\right) \qquad (30)$$

$$\mathbf{e}_{\mathbf{k},-1} = \frac{1}{\sqrt{2}}\left(i\mathbf{e}_{\mathbf{k},1} - \mathbf{e}_{\mathbf{k},2}\right) \qquad (31)$$

we have the following result

$$<\mathbf{J}_S> = \frac{\varepsilon_0}{2\omega}\bar{\mathbf{k}}\int d\mathbf{r}\left[|\mathbf{E}_{+1}(\mathbf{r})|^2 - |\mathbf{E}_{-1}(\mathbf{r})|^2\right] = \frac{\varepsilon_0}{\omega}\bar{\mathbf{k}}\int d\mathbf{r}\left[<|\mathbf{E}_{+1}(\mathbf{r},t)|^2> - <|\mathbf{E}_{-1}(\mathbf{r},t)|^2>\right] \qquad (32)$$

where $\bar{\mathbf{k}}$ is the unit vector in the direction of the wave vector $\mathbf{k}$ and $\mathbf{E}_{\pm 1}(\mathbf{r})$ are the spatially dependent fields of helicity $\lambda = \pm 1$. In the last passage of Eq. (32), the cycle average replaces the factor of one half and it is now manifest that the following classical result compares favorably with the quantum-mechanical result [pag. 491 in Ref. 2]

$$<\mathbf{J}_S> = \frac{\varepsilon_0}{\omega}\sum_{\mathbf{k},\lambda}\lambda\bar{\mathbf{k}}\int d\mathbf{r}\,<|E_{\mathbf{k},\lambda}(\mathbf{r},t)|^2> \qquad (33)$$

where $\varepsilon_0 \int d\mathbf{r} <|E_{\mathbf{k},\lambda}(\mathbf{r},t)|^2 > / \omega$ is the analogue of the $\hbar \hat{n}_{\mathbf{k},\lambda}$ term with $\hat{n}_{\mathbf{k},\lambda}$ the number operator. In general, we could proceed further by deepening the meaning of such a comparison. However, for the purpose of the current proof, we analyze the first term on the right-hand side of Eq. (32). This can be elaborated as follows

$$<\mathbf{J}_s>_{\lambda=+1} = \frac{\varepsilon_0}{\omega}\overline{\mathbf{k}} \int d\mathbf{r} <|\mathbf{E}_{+1}(\mathbf{r},t)|^2> = \frac{\varepsilon_0}{\omega}\overline{\mathbf{k}} \int d\mathbf{r} <|\mathbf{E}(\mathbf{r},t)|^2> \qquad (34)$$

where, thanks to the generality of Eq. (25) that defines $\mathbf{J}_s$, $\mathbf{E}(\mathbf{r},t)$ could be any electric field whose components are written in the basis of orthogonal linear polarizations $\mathbf{e}_{\mathbf{k},1}$ and $\mathbf{e}_{\mathbf{k},2}$. At this point, let us underline that the vector $<\mathbf{J}_s>_{\lambda=+1}$ is again a constant of the motion. This is guaranteed by the fact that both fields $\mathbf{E}_{+1}(\mathbf{r},t)$ and $\mathbf{E}(\mathbf{r},t)$ satisfy the wave equation and are related to their own vector potential through the time derivative. It means that the conservation law of Eq. (27) applies to $<\mathbf{J}_s>_{\lambda=+1}$ and, for this reason, we can conclude that the absolute value of $<\mathbf{J}_s>_{\lambda=+1}$ is a constant, or

$$|<\mathbf{J}_s>_{\lambda=+1}| = \frac{\varepsilon_0}{\omega} \int d\mathbf{r} <|\mathbf{E}(\mathbf{r},t)|^2> = constant \qquad (35)$$

The conclusion is very important, because we have open options on $\mathbf{E}(\mathbf{r},t)$. We can choose $\mathbf{E}(\mathbf{r},t)$ as the field we had at the beginning in Eq. (11) that corresponds to the field of a single-mode light and, in such an instance, we find

$$|< \mathbf{J}_S >_{\lambda=+1}| = \beta\left(n + \frac{1}{2}\right) \qquad (36)$$

that confirms a constant value for $\beta$ being $n$ a constant number. To conclude, $\beta$ is constant by virtue of the general validity of the conservation law for $\mathbf{J}_S$ of the free radiation field. Therefore, if $\beta$ is a constant and appears in the energy of Eq. (24) with the same role that $\hbar$ plays in the quantum-mechanical energy, we can safely assume that there is more than a hint of confidence in the suggestion that $\beta$ might be $\hbar$.

## 6. Conclusions

To sum up, we have found an alternative way that enables us to quantize the electromagnetic energy without the use of quantum-mechanical operators. The procedure relies on the fundamentals of the multipole expansion of the electromagnetic field. The final result is consistent with the quantization rule of $n+1/2$ energy quanta. In this scenario, the vacuum field emerges naturally as the field that carries one unit of energy associated with the fundamental harmonic. Finally, the conservation of the intrinsic angular momentum is useful to prove that the proportionality constant between the energy and the angular frequency might be identified with $\hbar$.

In the end, it seems that whether classical electrodynamics should be declared unfitted to quantize the free radiation field and hopelessly inappropriate for the concept of the vacuum field is a matter for debate.

**References**

[1] P. W. Milonni, The Quantum Vacuum, Academic Press, Boston, 1994.

[2] L. Mandel, E. Wolf, Optical Coherence and Quantum Optics, Cambridge University Press, Cambridge, 1995.

[3] A. Messiah, Quantum Mechanics, Dover Publications, New York, 1999.

[4] M. O. Scully, M. S. Zubairy, Quantum Optics, Cambridge University Press, Cambridge, 1999.

[5] R. Loudon, The Quantum Theory of Light, Oxford University Press, Oxford, 2000.

[6] W. P. Schleich, Quantum Optics in Phase Space, Wiley, Berlin, 2001.

[7] C. Gerry, P. Knight, Introductory Quantum Optics, Cambridge University Press, Cambridge, 2005.

[8] J. D. Jackson, Classical Electrodynamics, Wiley, New York, 1999.

[9] M. Born, E. Wolf, Principles of Optics, Pergamon Press, Oxford, 1989 (pp. 640-642).

[10] M. S. Howe, Acoustics of Fluid-Structure Interactions, Cambridge University Press, New York, 1998 (pp. 62-64).

[11] N. H. Sleep, K. Fujita, Principles of Geophysics, Blackwell Science, Malden, 1997 (pp. 487-490).